\documentstyle[aps,prd]{revtex}
\begin{document}
\draft
\title{A small universe after all?}
\author{ Neil J. Cornish \& David N. Spergel}
\address{Department of Astrophysical Sciences, Peyton hall,
Princeton University, Princeton, NJ 08544-1001, USA}

\twocolumn[\hsize\textwidth\columnwidth\hsize\csname
           @twocolumnfalse\endcsname

\maketitle
\widetext
\begin{abstract}
The cosmic microwave background radiation allows us to measure both the
geometry and topology of the universe. It has been argued that the COBE-DMR
data already rule out models that are multiply connected on scales smaller
than the particle horizon. Here we show the opposite is true: compact (small)
hyperbolic universes are favoured over their infinite counterparts. For
a density parameter of $\Omega_o=0.3$, the compact models are a better
fit to COBE-DMR (relative likelihood $\sim 20$) and the large-scale
structure data ($\sigma_8$ increases by $\sim 25$\%).
\end{abstract}
\bigskip
\medskip
]

\narrowtext


Measurements of the cosmic microwave background radiation (CMBR) provide
a powerful probe of the geometry and topology of the universe.
The geometry of the universe is reflected in the height, position and
spacing of acoustic peaks in the CMBR angular power
spectrum\cite{hu}, while the
topology of the universe is betrayed by matched circles in the
microwave sky\cite{css1,css2}. The topology of the universe also
influences the angular power spectrum by ``quantising'' the spectrum
of density fluctuations, and in many cases, by imposing a long
wavelength cut-off. An infrared cut-off in the spectrum of
density fluctuations translates into a
suppression of large angle CMBR fluctuations on the surface of last
scatter. This effect has been used to rule out
flat\cite{flat} and hyperbolic\cite{janna} models
with toroidal topologies.

Earlier we conjectured\cite{css3} that similar negative conclusions would
not apply to generic compact hyperbolic models as the majority of the
large angle CMBR power in a hyperbolic universe
comes not from the surface of last scatter, but from
the decay of curvature perturbations along the line of sight. To a lesser
extent, the same will be true in the newly popular flat models with a
cosmological constant, or in flat models with other forms of exotic dark
matter such as tangled string networks\cite{pen}. In what follows we shall
not only verify our conjecture, but show that finite hyperbolic
models actually provide a significantly better fit to the COBE-DMR data
than their infinite counterparts.

Our main focus will be on universes with compact hyperbolic spatial
sections\cite{sciam} as these are the most appealing from a theoretical
standpoint. However, the growing body of observational evidence favouring a
flat universe with a cosmological constant\cite{neta} prompts us to
reconsider models with three-torus topology. We refer to a model as being
``small'' if its comoving spatial volume is less than the comoving
volume enclosed by the particle horizon (measured in the covering space).
The ratio of the horizon volume to the volume of the space
exceeds 500 for several of the models we
looked at. Small universe models are obtained from the usual FRW models
by making identifications between different points in space. These
identifications break global isotropy and homogeneity, but do not alter
the evolution history. There is one caveat to the last statement: by
altering the mode spectrum, the topological identifications will alter
the vacuum structure, leading to a Casimir-like vacuum energy
that could alter the dynamics. If an effective cosmological constant
could be linked to the universe having non-trivial topology, we would
have a strong motivation for re-considering flat models.

On large angular scales, temperature fluctuations in the CMBR are
related to the fluctuations in the gauge-invariant gravitational
potential $\Phi$ by the Sachs-Wolfe equation
\begin{equation}\label{sw}
{\delta T(\theta, \phi) \over T}
= \frac{1}{3} \Phi(\eta_{sls}, r_{sls},\theta,\phi)
+2\int_{\eta_{sls}}^{\eta_{o}} \Phi'(\eta,
r,\theta,\phi)\,d\eta \, .
\end{equation}
Here $\eta$ denotes the conformal time, $\Phi'=\partial_\eta \Phi$,
 and $\eta_{sls}$ and $\eta_o$ denote recombination and the present
day respectively. The evolution of the gauge-invariant potential
from last scatter until today is described by
\begin{equation}\label{evol}
\Phi''+3{\cal H}(1+c_s^2)\Phi'-c_s^2\nabla \Phi + (2{\cal
H}'+(1+c_s^2) {\cal H})\Phi = 0 \, ,
\end{equation}
where ${\cal H}$ is the conformal Hubble factor
and $c_s$ is the sound speed in the cosmological fluid. In a flat matter
dominated universe we have $c_s=0$, ${\cal H}=2/\eta$ and
(\ref{evol}) tells us that $\Phi'=0$. Consequently, the second term in
(\ref{sw}) vanishes and the temperature fluctuations are all imprinted 
when matter and radiation decouple. The presence of either curvature
or a cosmological constant alters the time evolution of the
expansion rate ${\cal H}$, leading to a decay of the potential $\Phi$.
In these models, the line-of-sight integral (Integrated Sachs-Wolfe,
or ISW effect) in (\ref{sw}) can be the
dominant source of large angle temperature fluctuations.

The gravitational potential can be expanded in terms of the eigenmodes,
$\Psi_q$, of the Laplacian:
\begin{equation}\label{pot}
\Phi = \sum_{q} \sum_{n} \delta^{n}_{q} \, \Psi_q \, .
\end{equation}
In the above equation $n$ denotes the multiplicity of each eigenmode
and $\delta$ denotes the amplitude. The
eigenmodes are found as solutions of the equation
$\nabla \Psi_q = -q^2 \Psi_q$. The geometry of the space determines the
form of the Laplacian operator $\nabla$, while the topology determines
the boundary conditions. Though it is a simple exercise to write down the
eigenmodes for any of the 10 flat topologies, hyperbolic manifolds
have defied description. The first breakthrough
came last year when Inoue found the 14 lowest eigenmodes of the
Thurston space\cite{ino} using a numerical method\cite{as} originally
developed for 2-dimensional manifolds. By refining the method and using
a more powerful computer, Inoue extended the count to include the
first 36 eigenmodes. These modes have since been used to study the
CMBR\cite{ino2} in a universe with Thurston topology. Using the same
numerical method, Aurich\cite{ralf} has studied the CMBR in a
small hyperbolic orbifold with tetrahedral topology.
Recently a new, fully automated, algorithm for finding the
eigenmodes\cite{modes} was discovered and implemented. The list of
solutions has grown from 36 to several thousand in the past two weeks.
Our cosmological simulations use the first 100+ modes for each of 4 spaces
selected from the {\em SnapPea}\cite{weeks} census of compact
hyperbolic manifolds. Our selections are: the Weeks space, m003(-3,1), as
it is smallest know; the Thurston space, m003(-2,3), in order to compare
our results to Inoue's; and two larger examples, s718(1,1) and v3509(4,3),
to see how the size of the space influences the CMBR. In units of
the curvature radius cubed, the volumes of our selections are
0.9427, 0.9814, 2.2726 and 6.2392 respectively. To get a feel for
how the eigenmodes look, Figure 1 shows the lowest eigenmode of
the Weeks space extended across the entire Poincar\'{e} ball. The
Poincar\'{e} representation is a conformal mapping of $H^3$ into a
unit ball in $E^3$.

\begin{figure}[h]
\vspace{57mm}
\includegraphics{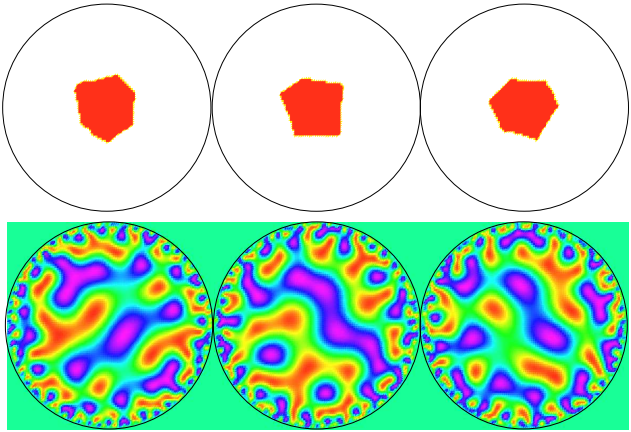}
\caption{The first eigenmode of the Weeks space.
The three views in each panel are, respectively,
the $x=0$, $y=0$ and $z=0$ slices through the Poincar\'{e} ball.
The upper panel shows the corresponding cross sections through the
fundamental cell.}
\end{figure}

The cosmological simulations are readily performed in spherical
coordinates, where the eigenmodes can be expanded:
\begin{equation}
\Psi_q = \sum_{\ell=0}^{\infty} \sum_{m=-\ell}^{\ell}
A_{q\ell m} R_{q\ell}(r) Y_{\ell m}(\theta, \phi) \, .
\end{equation}
Here the $Y_{\ell m}$'s are spherical harmonics and the
radial function $R_{q\ell}$ are either spherical (flat space) or
hyperspherical (hyperbolic space) Bessel functions. It is
worth emphasising how dramatically different the eigenmodes look
in compact flat spaces and compact hyperbolic spaces.
In flat space the expansion coefficients $A_{q \ell m}$ are 
peaked around certain values of $\ell$, and we can always rotate the
coordinate system so that $A_{q \ell m}=0$ for $m\neq 0$. In other
words, global isotropy is badly broken.
In contrast, the $A_{q \ell m}$'s for hyperbolic eigenmodes
are statistically independent of $\ell$ and $m$. The $A_{q \ell m}$'s
are gaussian distributed pseudo-random numbers with variance
proportional to $1/k^2$\cite{modes}. While it is possible to
assign a wavenumber $k=\sqrt{q^2-1}$ to hyperbolic eigenmodes, it is
impossible to assign a wavevector. They are essentially omni-directional.
You do not need inflation to explain why the
microwave sky is nearly isotropic in a small
hyperbolic universe\cite{css1}.

Once the eigenmodes have been found, it is a simple though laborious
task to evaluate equation (\ref{sw}) using equations
(\ref{evol}) and (\ref{pot}). We
took the mode amplitudes to be gaussian random numbers with variance
$\sqrt{k}/(k^2+4)$ (hyperbolic space) or $1/k^{3/2}$ (flat space).
These are the standard inflationary power
spectra.

\begin{figure}[h]
\vspace{70mm}
\includegraphics{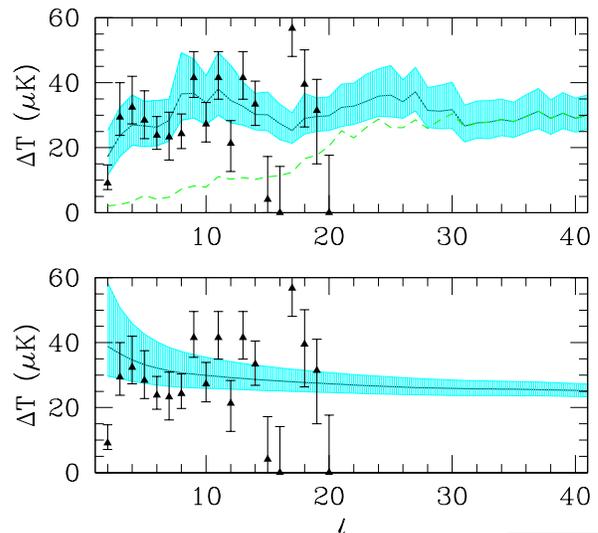}
\caption{The angular power spectrum $\Delta T=\sqrt{\ell
(\ell+1)C_\ell/6}$ for the Weeks universe (upper panel) and
the corresponding infinite universe (lower panel). Both models
have $\Omega_o=0.3$. The shaded areas mark the 1-$\sigma$ cosmic variance
intervals and the data points are from the 4-year COBE-DMR data. The dashed
line in the upper panel is the surface of last scatter contribution to the
angular power spectrum.}
\end{figure}

Figure 2 shows the angular power spectrum for a universe
with density parameter $\Omega_o=0.3$ and Weeks topology. Also
shown is the angular power spectrum for the corresponding infinite
model, along with the COBE-DMR data points\cite{kris}.
The effect of varying the density parameter is shown in Figure 3, while
the effect of varying the volume of the space is shown in Figure 4.
Notice that the large volume space v3509(4,3) produces an angular
power spectrum very similar to the infinite model shown in Figure 1.

We quantified how well each model fit the COBE-DMR data by modeling
the spread in the $\Delta T$'s by smooth probability distributions
(the statistics are not gaussian) and performing a standard likelihood
analysis. Our results are based on 1000 realizations of each topology
for five values of the density parameter. The likelihoods relative

\begin{figure}[h]
\vspace{70mm}
\includegraphics{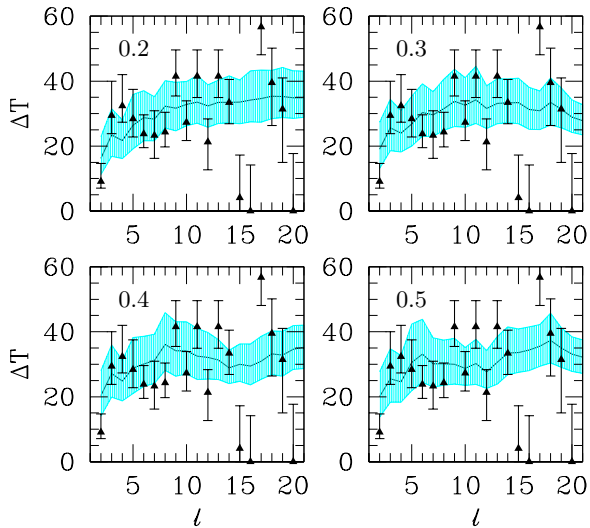}
\vspace{0mm}
\caption{Angular power spectra for the Thurston universe with
$\Omega_o=0.2$, 0.3, 0.4 and 0.5. These results compare
well with those found in Ref.~[12]}
\end{figure}
\vspace*{-2mm}
\begin{picture}(0,0)
\put(40,230){{\small 0.2}}
\put(145,230){{\small 0.3}}
\put(40,135){{\small 0.4}}
\put(145,135){{\small 0.5}}
\end{picture}

\begin{figure}[h]
\vspace{70mm}
\includegraphics{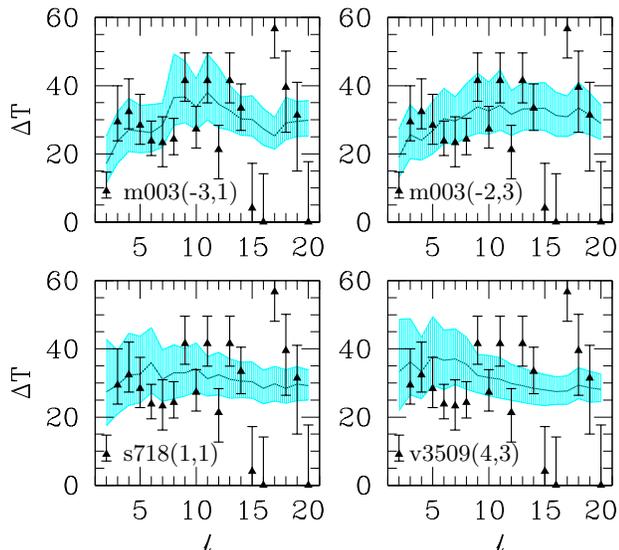}
\vspace{0mm}
\caption{Angular power spectra for 4 compact
hyperbolic universes with $\Omega_o=0.3$.}
\end{figure}
\vspace*{-2mm}
\begin{picture}(0,0)
\put(37,170){{\small m003(-3,1)}}
\put(145,170){{\small m003(-2,3)}}
\put(37,71){{\small s718(1,1)}}
\put(145,71){{\small v3509(4,3)}}
\end{picture}

\vspace*{-5mm}

\begin{table}
\caption{Relative likelihoods.}
\begin{tabular}{cccccccc}
& $\Omega_o$ & 0.2 & 0.3 & 0.4 & 0.5 & 0.6 & \\
\hline
& m003(-3,1) & 20.1 & 15.5 & 4.2 & 2.9  & 2.7 & \\
& m003(-2,3) & 30.3 & 25.4 & 7.5 & 4.1  & 5.0 & \\
& s718(1,1)  & 21.5 & 8.2  & 3.7 & 2.1  & 1.5 & \\
& v3509(4,3) & 12.2 & 2.2  & 1.5 & 2.5  & 1.1 & \\
& infinite   & 0.14 & 0.26 & 0.52 & 0.92 & 1.2 & \\
\end{tabular}
\end{table}

\noindent to the corresponding infinite model
are listed in Table~1, along with the likelihoods of the infinite
hyperbolic models relative to a fiducial flat matter dominated universe.

The lower the density, the better the compact models fare relative
to their infinite counterparts. The mechanism behind this result can
be seen at work in Figure~1. As we decrease the density, the infrared
cut-off in the mode spectrum reduces the contribution from the first
term in equation (\ref{sw}), while the line of
sight contribution from the second term becomes increasingly important.
Rather miraculously, the two effects almost precisely cancel out for
the low volume models, resulting in a flat or mildly tilted spectrum.
On small scales there is no difference in the shape of the angular
power spectra for compact and infinite models, but the compact models
have a higher COBE-DMR normalization. This helps raise the predicted
size of density fluctuations on 8 $h^{-1}$ Mpc scales from the
low value of $\sigma_8 =0.6$ for an infinite model,
to the larger value of $\sigma_8 =0.75$ for the Weeks model (both
with $\Omega_o=0.3$). Measurements of the present day cluster
abundance\cite{pen} favour $\sigma_8 = 0.9\pm 0.1$ if $\Omega_o=0.3$.
At this stage we should stress that our results primarily affect the fit
to the COBE-DMR data. On small scales there is a growing body of
observational evidence\cite{amber} for
an acoustic peak at $\ell \sim 220$, which is consistent
with the universe being flat, not hyperbolic. We should also mention
that our results disagree with those of Bond {\it et al}\cite{dick}
based on the method of images, and agree with those of
Inoue {\it et al}\cite{ino2} based on the finite element method.

\
\begin{figure}[h]
\vspace{60mm}
\includegraphics{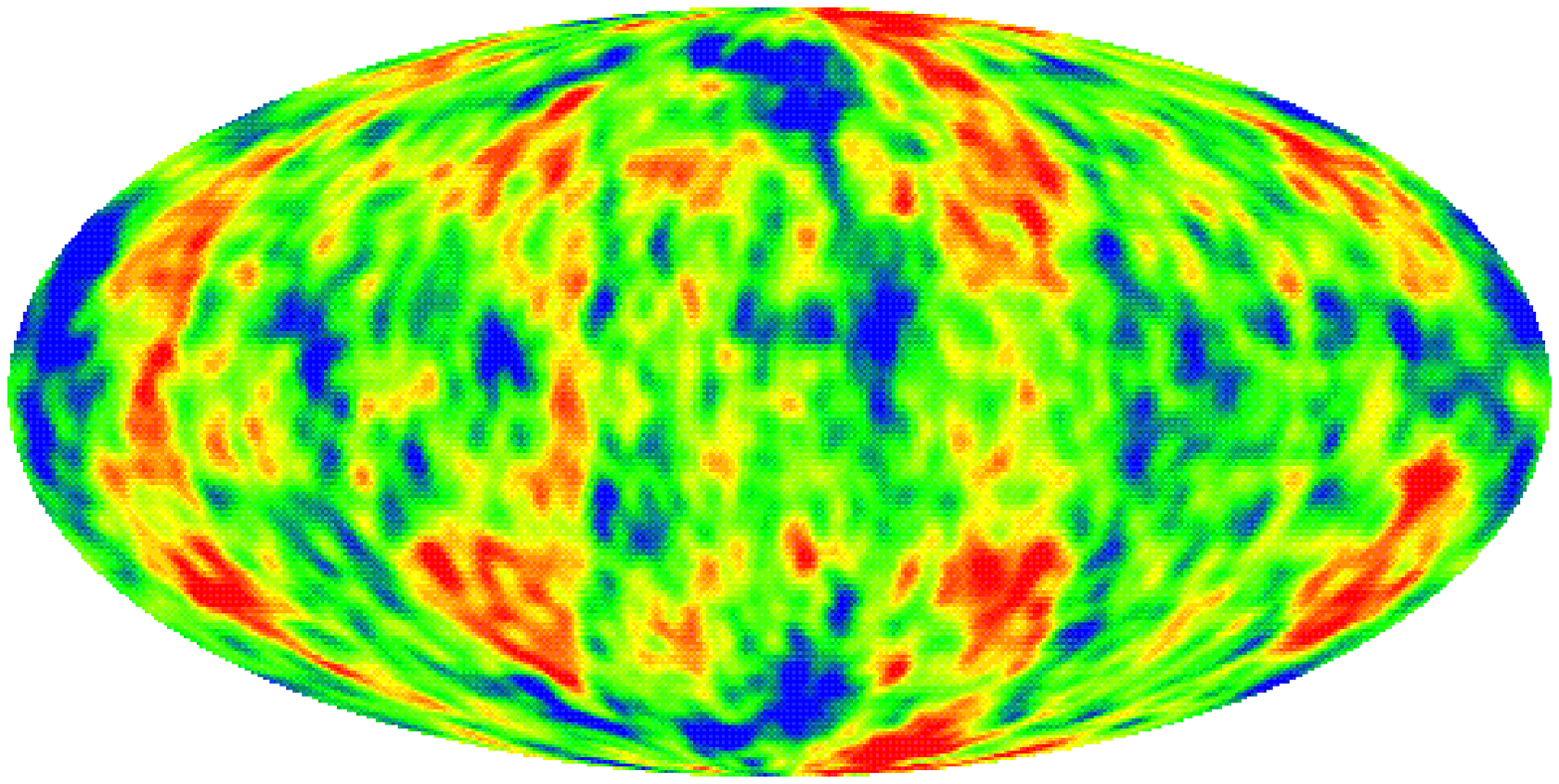}
\includegraphics{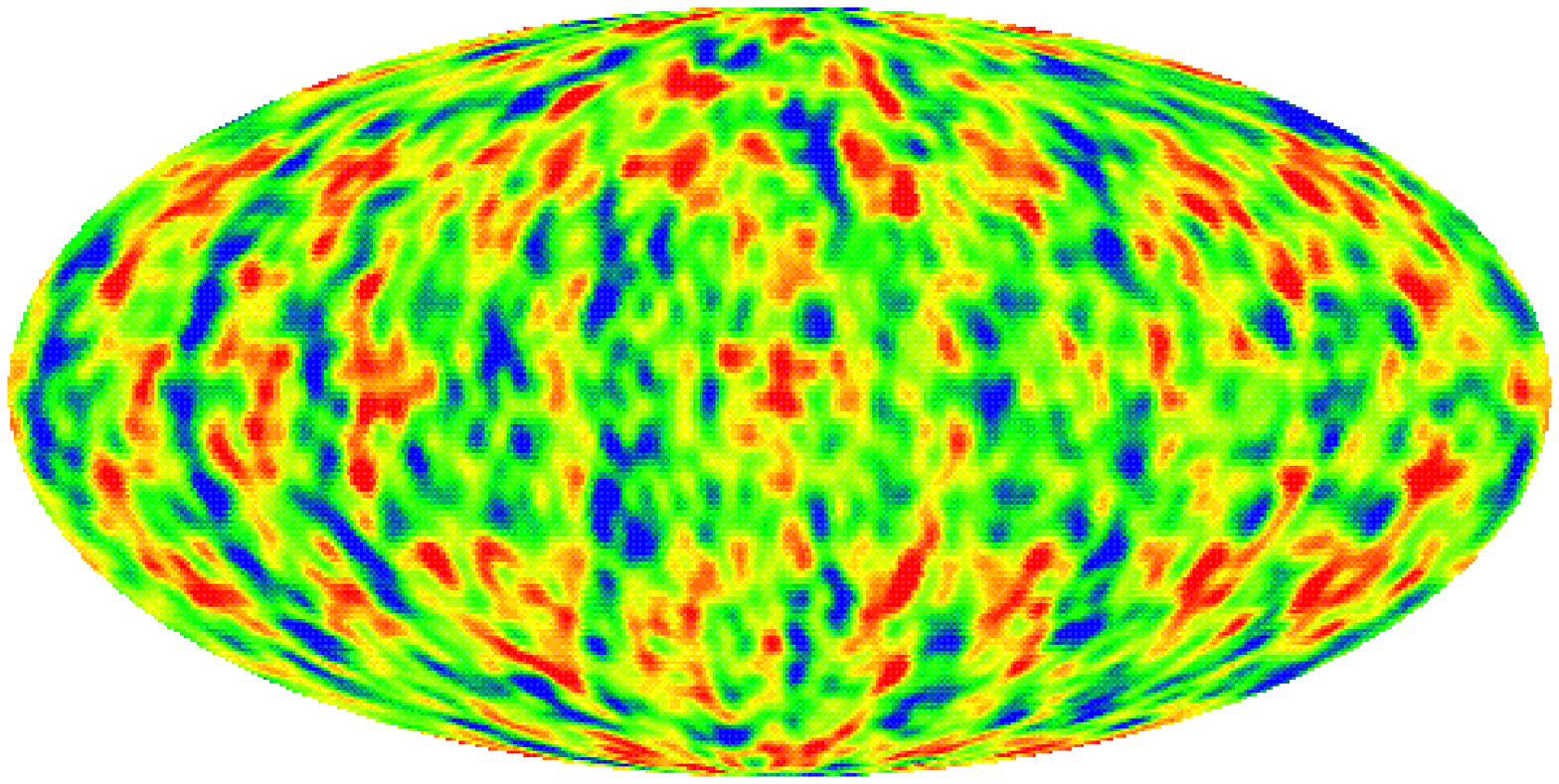}
\vspace{8mm}
\caption{The microwave sky in a flat three torus universe, with (upper
panel) and without (lower panel) a cosmological constant term.}
\end{figure}

If the universe is flat, but dominated by a cosmological constant, could
the toroidal universe models be saved by contributions
from the the line of sight integral in (\ref{sw})? Our preliminary
investigation suggests that they may, but not so much in the
power spectrum department as in the breaking of global isotropy.
For example, a flat cubical three-torus with side length equal to
half the horizon distance has a relative
likelihood of $0.26$ when $\Omega_\Lambda=0$ and $0.25$ when
$\Omega_\Lambda=0.7$. The real difference between these two models
can be seen in Figure~5. Because the ISW effect mixes
together different modes sampled at different points, it helps to
hide the egregious breaking of global isotropy that lead to the
matter dominated versions of these models being ruled out\cite{flat}.

Returning to the compact hyperbolic models, we want to see if the
ISW effect ruins the the matched circle test\cite{css1} 
for non-trivial topology. The matched ``circles
in the sky'' occur wherever the surface of last scatter self-intersects.
Since the surface of last scatter is a 2-sphere, the intersections
occur along circles. We see two copies of each circle of intersection,
centered at different points on the sky. The portion of the microwave
temperature coming from the surface of last scatter will be identical
around each circle. However, the ISW contribution will be uncorrelated.

\
\begin{figure}[h]
\vspace{35mm}
\includegraphics{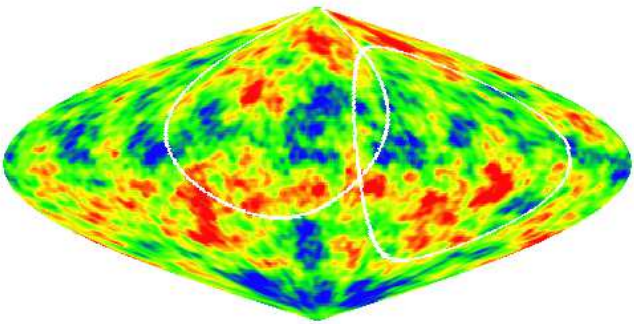}
\vspace{8mm}
\caption{The microwave sky in one realization of the Weeks universe with
$\Omega_o=0.3$. The white lines mark one pair of matched circles.}
\end{figure}

\begin{figure}[h]
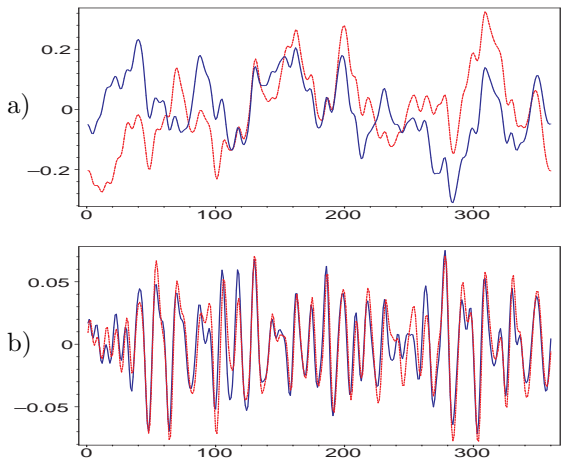

\vspace{60mm}
\includegraphics{weeks03circ5.ps}
\includegraphics{weeks03circ5f.ps}
\caption{The temperatures around a pair of matched circles before
and after the low multipoles are removed.}
\end{figure} 

\vspace*{-2mm}
\begin{picture}(0,0)
\put(1,173){a)}
\put(1,83){b)}
\end{picture}

Taking one realization of the Weeks universe (Figure~6), we find the
temperatures around a pair of matched circles (Figure~7a) and
see that the match is poor. However, the ISW effect only operates on
large angular scales, so we filter Figure~6 to remove
all power below $\ell=21$. The temperature match for the
filtered sky is shown in Figure~7b. The correlation
coefficient\cite{css2} increases from $0.29$ to $0.92$ after
filtering out modes with $\ell \leq 20$. The matched circle pairs will
persist until the visible universe is simply connected, which occurs
at around $\Omega_o=0.95$ for most models in the {\em SnapPea} census.
If we do live in a small universe, the Microwave Anisotropy Probe
will find matched circles in the sky when it starts collecting
data in 2001.

\bigskip

This work grew out of earlier collaborations and extensive discussion
with Glenn Starkman and Jeff Weeks. Financial support was provided by
NASA through their funding of the Microwave Anisotropy Probe
satellite mission {\tt http://map.gsfc.nasa.gov/}.


\begin{references}
\bibitem{hu} A.G. Doroshkevich, Ya. B. Zel'dovich \& R.A. Sunyaev,
Sov. Astron. {\bf 22}, 523 (1978); W. Hu and M. White,
Astrophys. J. {\bf 471}, 30 (1996). 
\bibitem{css1} N.J. Cornish, D.N. Spergel \& G. Starkman,
Phys. Rev. Lett. {\bf 77}, 215 (1996).
\bibitem{css2} N.J. Cornish, D.N. Spergel \& G. Starkman, Class. Quant.
Grav. {\bf 15}, 2657 (1998).
\bibitem{flat} I.Y. Sokolov, JETP Lett. {\bf 57}, 617 (1993);
A.A. Starobinsky, JETP Lett. {\bf 57}, 622 (1993);
D. Stevens, D. Scott \& J. Silk,
Phys. Rev. Lett. {\bf 71}, 20 (1993);
A. de Oliveira Costa \& G.F. Smoot,
Ap. J. {\bf 448}, 477 (1995); A. de Oliveira Costa,
G.F. Smoot \& A.A. Starobinsky Ap. J. {\bf 468}, 457 (1996).
\bibitem{janna} J.J. Levin, J.D. Barrow, E.F. Bunn \& J. Silk,
Phys. Rev. Lett. {\bf 79}, 974 (1997).
\bibitem{css3} N.J. Cornish, D. Spergel \& G. Starkman,
Phys. Rev. D{\bf 57}, 5982 (1998).
\bibitem{pen} D.N. Spergel, U. Pen, Astrophys. J. {\bf 491}, L67 (1997).
\bibitem{sciam} For popular introductions see J.P. Luminet, G.D. Starkman
\& J.R. Weeks, Sci. Am., April (1999).
\bibitem{neta} For a review see N.A. Bahcall, J.P. Ostriker,
S. Perlmutter \& P.J. Steinhardt, Science {\bf 284}, 1481 (1999).
\bibitem{ino} K.T. Inoue, astro-ph/9810034, (1998).
\bibitem{as} R. Aurich and F. Steiner, Physica D{\bf 64}, 185 (1993).
\bibitem{ino2} K.T. Inoue, K. Tomita \& N. Sugiyama, astro-ph/9906304 (1999).
\bibitem{ralf} R. Aurich, astro-ph/9903032, (1999).
\bibitem{modes} N.J. Cornish \& D.N. Spergel, math.DG/9906017 (1999).
\bibitem{weeks} J. Weeks, {\em SnapPea}: http://www.northnet.org/weeks.
\bibitem{jw} J.R. Weeks, Class. Quant. Grav. {\bf 15}, 2599 (1998).
\bibitem{kris} K.M. G\'{o}rski, 31$^{\rm st}$ Rencontres de Moriond,
Eds. F.R. Bouchet {\it et al}, (1997); C.L. Bennett {\it et al}, 
Astrophys. J. {\bf 464}, L1 (1996).
\bibitem{pen} U.-L. Pen Astrophys. J. {\bf 498}, 60 (1998).
\bibitem{amber} A. Miller {\it et al}, submitted to Ap. J. (1999).
\bibitem{dick} J.R. Bond, D. Pogosyan \& T. Souradeep,  Class. Quant.
Grav. {\bf 15}, 2671 (1998).

\end{references}
\end{document}